\documentclass[preprint,1p,times]{elsarticle}

\usepackage[utf8]{inputenc}
\usepackage{amsmath,amsfonts}
\usepackage{mathtools}
\DeclarePairedDelimiter\floor{\lfloor}{\rfloor}
\usepackage{graphicx}
\usepackage{textcomp}
\usepackage[small,bf,hypcap=true]{caption}
\usepackage{color}
\usepackage[shortlabels]{enumitem}
\usepackage{nameref}
\usepackage[english]{babel}

\begin{document}

\title{Pattern selection in a ring of Kuramoto oscillators}

\author[bbte]{K\'aroly D\'enes}

\author[bbte,gth]{Bulcs\'u S\'andor}

\author[bbte]{Zolt\'an N\'eda\corref{cor1}}
\ead{zoltan.neda@phys.ubbcluj.ro}

\cortext[cor1]{Corresponding author}

\address[bbte]{Babe\c{s}-Bolyai University, Department of Physics, 1 Kogălniceanu str., 400084 Cluj, Romania}
\address[gth]{Goethe University Frankfurt, Institute for Theoretical Physics, Max-von-Laue-Str. 1, 60438 Frankfurt am Main, Germany}

\begin{abstract}
Emergence of generalized synchronization patterns in a ring of identical and locally coupled Kuramoto-type rotators are investigated by different methods. 
These approaches offer
a useful visual picture for understanding the complexity of the dynamics in the high dimensional state-space of this system. 
Beside the known stable stationary points  novel unstable states are revealed. 
We find that the prediction of the final stationary state is limited by the presence of such saddle points. 
This is illustrated by considering and comparing two different attempts for forecasting the final stationary state. 
\end{abstract}

\maketitle

\section{Introduction}

Collective behavior in ensembles of interacting oscillators is one of the oldest problems in the field of dynamical systems
and statistical physics \cite{Strogatz2001,pikovski2001}. Interestingly however, this field is still active, raising new problems \cite{panaggio}, revealing further surprises \cite{yeldesbay} and 
offering applications and modeling tools for many other areas of science and engineering \cite{wood,ott}.  

Synchronization of non-identical and coupled oscillators is an intriguing fact observed in many real systems. The Kuramoto model  
\cite{Kuramoto} is probably the most  widely studied system for modeling such synchronization phenomena. 
For globally coupled rotators it exhibits an order-disorder transition, which is useful to explain emerging synchronization in 
physical, social or biological systems \cite{Acebron}.  By varying the interaction topology among the rotators, many variants of the original model were 
studied analytically and numerically.  It was found that the topology of the interaction determines 
the nature of the emerging collective behavior. In such sense the Kuramoto model was considered both on regular and random graphs \cite{mirollo1988a,mirollo1988b,coutinho} using interactions between neighbors of different order \cite{lumer,wiley_strogatz_2006}. The model was generalized 
also by considering a mixture of attractive and repulsive couplings \cite{daido}. For locally coupled nonidentical oscillators a rich variety of collective behaviors was found: 
frequency locking, phase synchronization, partial synchronization or incoherence.
Time-delay in the interactions between the active neighbors introduces an extra complexity in the Kuramoto model by drastically increasing 
its dimensionality \cite{Strogatzdelay,earl,Acebron,panaggio}. 
It yields also new surprises in large oscillator ensembles, by generating novel states where some of the 
oscillators are synchronized while the others remain completely disorganized. 
Such states are called "chimera" states, and they have been observed in many different coupling topologies both in systems with time-delay 
\cite{sethia,panaggio,yeldesbay} or in the absence of time-delay \cite{kuramoto2002coexistence,schmidt2015cluster,
nicolaou,Smirnov2017}. 
In the present work due to the simplified coupling topology (nearest neighbor coupling) and the absence of time-delay, chimera-like states are not 
expected to appear \cite{wiley_strogatz_2006}.  

The system investigated by us consists of $N$ classical Kuramoto oscillators displayed in a ring-like topology,
with identical intrinsic frequencies $\omega_0$, each oscillator being coupled to its nearest neighbors with coupling strength $K$. 
The dynamics of the system is given by the coupled first order differential equation system:
\begin{equation}
 \label{eq.1}
 \dot{\theta_i}=\omega_0+K [\sin(\theta_{i-1}-\theta_i)+\sin(\theta_{i+1}-\theta_i)]\,,
\end{equation}
with $\theta_i=\theta_i(t)$ being the time-dependent phase of the $i$-th oscillator, $i=\overline{1,N}$. Periodic boundary conditions $\theta_{N+1}=\theta_1$ and $\theta_0=\theta_{N}$ are assumed. 
The symmetry of the system allows many dynamically stable stationary states (different types of collective behavior)  to appear \cite{wiley_strogatz_2006,Roy2012888,diaz,ochab}. Such states are generalized synchronization states in form of self-closing traveling waves with a fixed winding number,~$m$, all the oscillators having the same $\omega_0$ frequency. The simplest and most probable state is the classical synchrony ($m=0$)
where all rotators move in phase.  The other stable stationary states ($m=\pm1,\pm 2,\ldots$) are characterized by a locked phase shift between the neighbors. 

A detailed mathematical study for a generalised version of the model (\ref{eq.1}) was very recently considered by Burylko et. al \cite{burylko}. In this study one allows also interactions with neighbors of higher rank, and the fixpoints are thoroughly analyzed.     
The present study is limited however only on nearest-neighbor interactions. We fix the $T_0\equiv2 \pi/\omega_0=\pi$ natural period for the oscillators (defining by this the time-unit), 
and study the dynamics of the system in the view of the predictability of the final stationary state.  

\section{Emergent patterns}
\label{sec:overwiev}

First we give an overview in an original approach of the known results for the stationary states and introduce also some basic 
concepts and notations that are used in the following sections.  

\subsection{Stationary states}
\label{sec:stationary}

For convenience reasons we switch to a reference 
frame rotating with the natural frequencies of the oscillators:
\begin{equation}
 u_i(t)=\theta_i(t)-\omega_0 t\,.
 \label{eq:variables_u}
\end{equation}
The equation of motion in the rotating frame will be:
\begin{equation}
 \label{eq.stab.2}
 \dot u_i = K[\sin(u_{i-1}-u_i)  
 +\sin(u_{i+1}-u_i)] = F(u_{i-1},u_i,u_{i+1})\,.
\end{equation}
In this reference frame Eq.~(\ref{eq.stab.2}) is a gradient system having the following potential function \cite{wiley_strogatz_2006}:
\begin{equation}
 \label{eq.potential}
 V=-\frac{K}{2}\sum_{j=1}^N\left(\cos(u_{j-1}-u_j)+\cos(u_{j+1}-u_j)\right).
\end{equation}
Eq.~(\ref{eq.stab.2}) is now equivalent to $\dot u_i=-\partial V /\partial u_i$. Being a gradient system means that the stationary states are always fixpoints while limit cycles or any other kind of 
attractors are not allowed \cite{Hirsch2013}. These stationary states correspond to local minima, maxima or saddle points of~$V$.

Fixpoints require $\dot u_i=0$, hence:
\begin{equation}
 \label{eq.fixpoint}
 K[\sin(u_{i-1}-u_i)+\sin(u_{i+1}-u_i)] =0.
\end{equation}
Converting this sum to a product will yield:
\begin{equation}
 \label{eq.fixpoint.product}
 K\sin\left(\frac{u_{i+1}-2u_i+u_{i-1}}{2}\right)\cos\left(\frac{u_{i+1}-u_{i-1}}{2}\right)=0.
\end{equation}
If either of the two trigonometric functions evaluates to 0 we have stationarity. Thus the two conditions are:
\begin{equation}
 \label{eq.cond1}
 u_{i+1}-2u_i+u_{i-1}=2k_i \pi,\quad k_i \in\mathbb{Z},
\end{equation}
and
\begin{equation}
 \label{eq.cond2}
 u_{i+1}-u_{i-1}=(2q_i+1)\pi, \quad q_i\in\mathbb{Z}.
\end{equation}
The above conditions can be satisfied for the whole system in the following manners: 
\begin{enumerate}[(a)]
\item  condition (\ref{eq.cond1}) is fulfilled for  all $i$ indeces, 
\item condition (\ref{eq.cond2}) is true all over the system, 
\item for some $i$ values condition (\ref{eq.cond1}) holds, while for the other $i$ values condition (\ref{eq.cond2}) is true.
\end{enumerate}
While cases (a) and (b) conserve the symmetry of the system, case (c) will violate it, corresponding to a nontrivial symmetry breaking. 

\begin{figure}[t]
  \centering
  \includegraphics[width=0.5\linewidth]{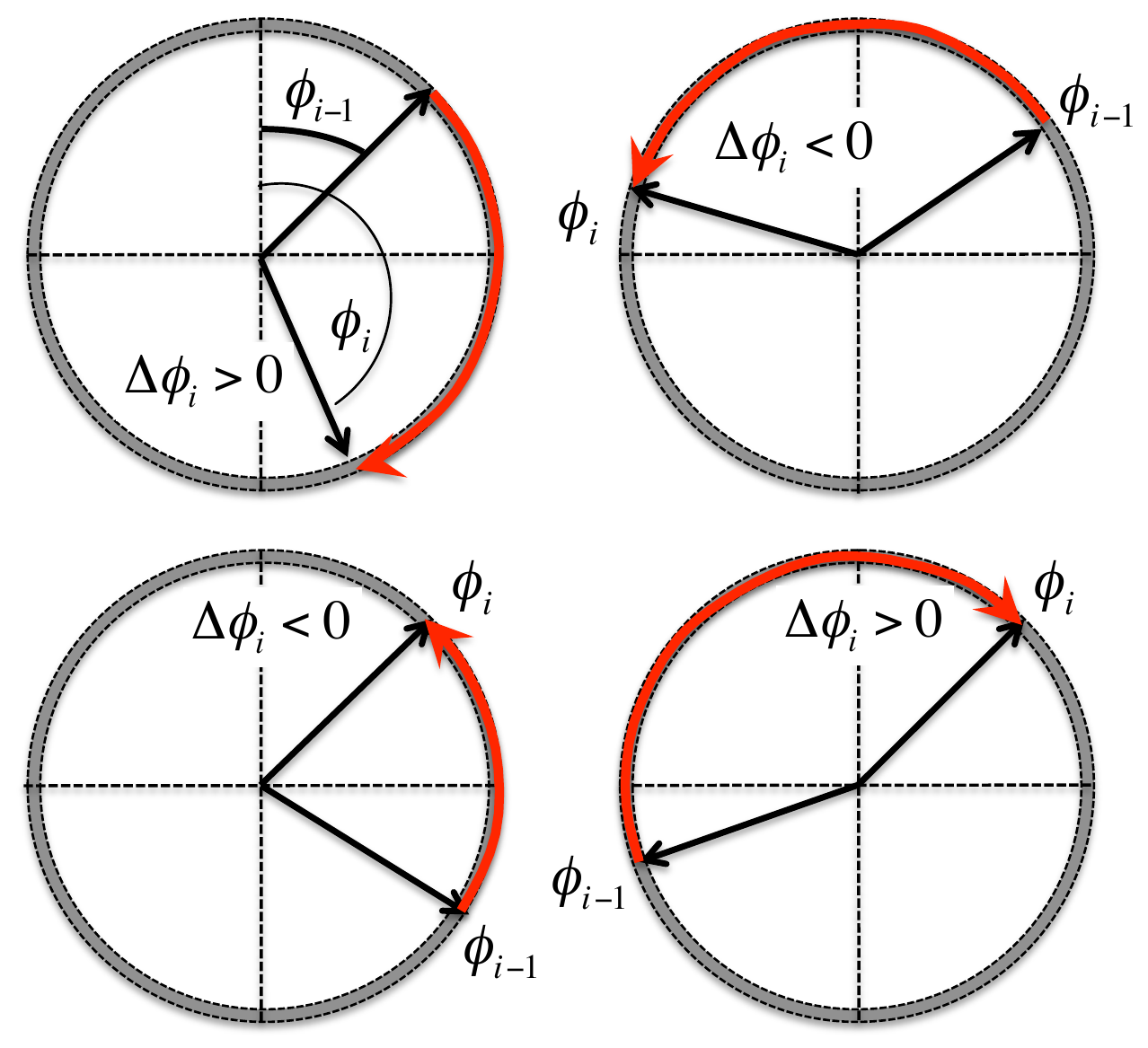}
  \caption{(color online) Illustration of the $\Delta \phi_i$ phase shifts, see also Eqs.~(\ref{deltafieq}).}
  \label{deltafi}
\end{figure}

To represent the state of the system on a unit circle it is convenient  to use a new phase variable: $\phi_i$ ($0\le\phi_i<2\pi$):
\begin{equation}
\phi_i=u_i\mod  2 \pi\,.
\end{equation}
Taking into account that the asymptotic solutions are characterized with fixed $u_i$ values, these are phase-locked states. Therefore the relative positions of oscillators $i$ and $i-1$ on the unit circle has to be characterized with a parameter 
$\Delta \phi_i$, named hereafter as {\em phase shift} between oscillators $i$ and $i-1$, which takes values between $-\pi$ and $\pi$ as it is illustrated in Fig.~\ref{deltafi}. 
In order to achieve this, the  $\Delta \phi_i$ parameter has to be defined as: 
\begin{equation}
\begin{aligned}
  \Delta \phi_i &= \phi_i-\phi_{i-1}  \quad  &&\textrm{for} \quad \phantom{2}-\pi \le \phi_i-\phi_{i-1} < \pi \\
 \Delta \phi_i &= \phi_i-\phi_{i-1} -2\pi  \quad  &&\textrm{for} \quad \phantom{22-}\pi \le \phi_i-\phi_{i-1} < 2\pi \\  
\Delta \phi_i &= \phi_i-\phi_{i-1} +2\pi  \quad  &&\textrm{for} \quad -2 \pi < \phi_i-\phi_{i-1} < - \pi\,,
\label{deltafieq}
\end{aligned}
\end{equation}

which can be written in a compact form by using the floor function ($f(x)=\floor{(x)}$):
\begin{equation}
 \Delta \phi_i=(\phi_i-\phi_{i-1})-2\pi \floor*{\frac{\phi_i-\phi_{i-1}+\pi }{2 \pi}}.
\label{deltafirel}
\end{equation}

Let us consider now $\Delta_N=\sum_{i=1}^{N} \Delta \phi_i$.
Using Eq.~(\ref{deltafirel}), we can write 
\begin{equation}
\Delta_N=\sum_{i=1}^N (\phi_i-\phi_{i-1})-2\pi\sum_{i=1}^N \floor*{\frac{\phi_i-\phi_{i-1}+\pi}{2\pi}}\,.
\label{deltaN1}
\end{equation}
In our ring topology $\phi_{0}=\phi_{N}$, so the first sum in Eq.~(\ref{deltaN1}) is zero. Since the terms in the second sum are all $0$ or $\pm1$, it results that the second sum should be an integer, $m\in\mathbb{Z}$. 
As a consequence, the sum of the phase shifts satisfy the relation (see also \cite{Roy2012888}):
\begin{equation}
\Delta_N=2\pi m,
\label{winding}
\end{equation}
where negative values of $m$ are also allowed. This relation has nothing to do with the dynamical equations of the system, it is solely a consequence of the imposed topology.

\begin{figure}
 \centering
 \includegraphics[width=0.75\linewidth]{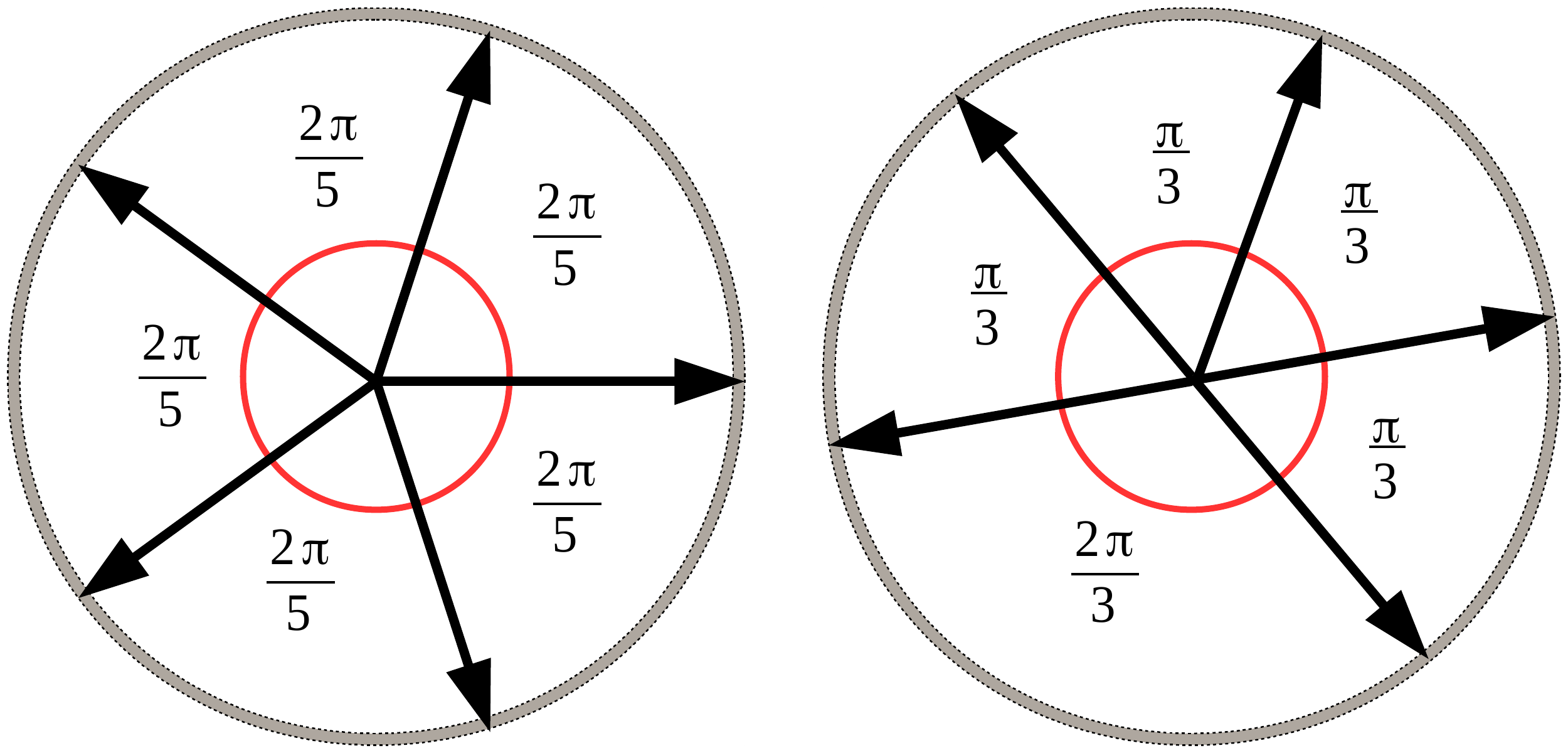}
 \caption{(color online) Left: Illustration of case (a) stationary state with $m=+1$ winding number for $N=5$ oscillators. All triplets are balanced, therefore the phase difference is constant over the system. Right: Illustration of a case (c) stationary state. This example depicts a system with $N=5$ oscillators with $m=+1$ winding number. One can immediately notice the broken symmetry
 of the system, since not all phase shifts are equal even though the $N$ and the winding number is the same as on the left. This indicates that there are pairs/triplets for which Eq.~(\ref{eq.cond2}) holds among the other balanced triplets determined by
 Eq.~(\ref{eq.cond1}). }
 \label{fig.case_c}
\end{figure}

Considering {\bf case (a)} for the fixpoint condition, we rewrite equations (\ref{eq.cond1}) using the $\phi_i$ variables:
\begin{equation}
 \label{eq.cond3}
 \phi_{i+1}-2\phi_i+\phi_{i-1}=2k_i \pi,\quad k_i \in\mathbb{Z}\,.
\end{equation}
Regrouping the multiples of $2\pi$ one can write the above condition also in terms 
of the phase shifts:
\begin{equation}
 \label{eq.cond5}
 \Delta\phi_{i+1}-\Delta\phi_{i}=2l_i\pi,\quad l_i \in\mathbb{Z}\,.
 \end{equation}
Since $\Delta\phi_i \in[-\pi,\pi)$ this condition is fulfilled only for $l_i=l=0$. Hence: 
\begin{equation}
 \label{eq.cond6}
 \Delta\phi_{i+1}=\Delta\phi_{i}\,.
\end{equation}
Consequently,
in these stationary states the phase shift $\Delta \phi_i =\Delta \phi$ is constant for all oscillator pairs (see Fig.~\ref{fig.case_c} left panel). This leads to the result:
\begin{equation}
 \label{eq.4}
 \Delta\phi=\frac{\Delta_N}{N}=2\frac{m}{N}\pi\,.
\end{equation}
Due to the fact that the phase shift satisfies $-\pi \le \Delta \phi < \pi$, we get that $-N/2 \le m < N/2$, with $m\in\mathbb{Z}$. 

Stationary states stemming from Eq.~(\ref{eq.cond6}) are defined thus by the $m$ number, referred from now on as the {\em state index} or {\em winding number}. Synchrony in the classical sense corresponds to the $0$ index, the other states being indexed from~$-N/2$ up to~$N/2-1$. As an example the case of $m=+1$ for $N=5$ oscillators is sketched in the left panel of Fig.~\ref{fig.case_c}.
The existence of these stationary states is well-known in the literature \cite{dodla,wiley_strogatz_2006,Roy2012888}, however other authors use different arguments to arrive at this result.

The other branch of stationary states given by {\bf case~(b)} is obtained by satisfying the condition in equations (\ref{eq.cond2}). 
With the same reasoning as in case (a) one arrives to:
\begin{equation}
 \label{eq.cond7}
 \Delta\phi_{i+1}+\Delta\phi_i=(2p_i+1)\pi, \quad p_i \in\mathbb{Z}\,.
 \end{equation}
Taking into account that $\Delta\phi_{i}\in[-\pi,\pi)$ the two phase shifts can only add up to $\pm\pi$ which is equivalent to $p_i\in\{-1,0\}$. Generally the $p_i$ parameter may be different for the pairs of rotators,
however it can be shown that it is constant over the whole system. 
In order to realize this, let us assume that the $p_i$ value changes for two consecutive pairs:
\begin{eqnarray}
\nonumber
\label{eq.cond8}
  \Delta\phi_{i+2}+\Delta\phi_{i+1}=\pm\pi\\
  \Delta\phi_{i+1}+\Delta\phi_{i}=\mp\pi.
\end{eqnarray}
Subtracting the two equations we get:
\begin{equation}
 \Delta\phi_{i+2}-\Delta\phi_{i}=\pm2\pi.
\end{equation}
This condition cannot be fulfilled since $-\pi \le \Delta \phi_i < \pi$ so the difference between two phase shifts is always greater than $-2\pi$ and smaller than $2\pi$, therefore $p_i=p_{i+1}=p$ ($p\in\{-1,0\}$) for all pairs. With this argument Eq.~(\ref{eq.cond7}) can be rewritten as:
\begin{equation}
\label{eq.cond.pm_b}
 \Delta\phi_{i+1}=\pm\pi-\Delta\phi_i\,.
\end{equation}

To gain some more
information about this type of states we sum up all the equations in (\ref{eq.cond7}):
\begin{equation}
 \sum_{i=1}^{N}\left ( \Delta\phi_{i+1}+\Delta\phi_i \right )=\sum_{i=1}^{N}(2p+1)\pi.
\end{equation}
Invoking the periodic boundary condition we can write:
\begin{equation}
 2\sum_{i=1}^{N}\Delta\phi_i=2\cdot2 m \pi=N(2p+1)\pi.
\end{equation}
Finally, by using $p\in\{-1,0\}$ we determine the possible values of the $m$ winding number:
\begin{equation}
 \label{stat.state.2}
 m=\pm \frac{N}{4}.
\end{equation}
These kind of states are only possible if $N$ is divisible by~4 and their number is infinite since there are infinite phase shifts for which the $\Delta\phi_{i+1}=\pm\pi-\Delta\phi_i$ condition
holds.

The symmetry violating {\bf case (c)} is a combination of conditions (\ref{eq.cond6}) for some $i$ values and (\ref{eq.cond7})
 with $p_i\in\{-1,0\}$, for the other $i$ indices. Since this is a highly unusual case, we present an example of such a nontrivial configuration for $N=5$ oscillators in the right panel of Fig. \ref{fig.case_c}.
 One can immediately realize that there are many possibilities to fulfill case (c). Following a similar argument as the one used in case (b) one can show that if there are triplets of oscillators satisfying condition (\ref{eq.cond7}) all of them must have the same $p_i=p$ ($p_i\in\{-1,0\}$) value. 
 As a consequence of these, all $\Delta \phi_i$ values are either a constant $\Delta \phi$ or 
 $(2p+1) \pi -\Delta \phi$. Assuming that there are $n$ number of $\Delta \phi_i=\Delta \phi$ phase shifts and consequently 
 $N-n$ phase shift with $(2p+1) \pi -\Delta \phi$ value, the imposed boundary condition (\ref{winding}) leads to
 \begin{equation}
 \label{eq.case_c.sum}
 \sum_{i=1}^N \Delta \phi_i=n\,\Delta \phi + (N-n)\,[(2p+1)\pi-\Delta \phi]=2m\pi, 
 \end{equation}
where $-N/2 \le m < N/2$ with $m\in\mathbb{Z}$.
 
\subsection{Stability of the stationary states}
\subsubsection{Case (a) states}
In {\bf case (a)} when condition (\ref{eq.cond1}) is fulfilled for  all $i$ indeces using the condition in (\ref{eq.4}) we identified the possible stationary states characterized by equal phase shifts $\Delta \phi =2 m \pi/N$ with  $-N/2 \le m < N/2$.
We analyze now their stability.
For this purpose we use the standard linearization near the equilibrium point.

The Jacobian of the system, evaluated at the equilibrium solution $\mathbf{u}^*=(u_1^*,\dots,u_i^*,\dots)$ is constructed as follows:
\begin{equation}
 \begin{split}
  J_{ij}=\left.\frac{\partial F(u_{i-1},u_i,u_{i+1})}{\partial u_j}\right|_{\mathbf{u}^*}= 
  K\Big[\cos(u^*_{i-1}-u^*_i)\delta_{i-1,j}-\\
 -\Big(\cos(u^*_{i-1}-u^*_i)+
 \cos(u^*_{i+1}-u^*_i)\Big)\delta_{i,j}+\\
 +\cos(u^*_{i+1}-u^*_i)\delta_{i+1,j}\Big]\,.
 \end{split}
 \label{eq:jacobian_general}
\end{equation}
In equilibrium $u^*_i-u^*_{i-1}=\Delta\phi$. Hence, we can write the 
Jacobian explicitly in the form of a circulant matrix:
\begin{equation}
\begin{split}
 \mathbf{J}=K \cos\Delta\phi
 \begin{pmatrix}
-2	&	1	&	0	&	\dots	&	0	&	0	&	1\\
1	&	-2	&	1	&	\dots	&	0	&	0	&	0\\
0	&	1	&	-2	&	\dots	&	0	&	0	&	0\\
\vdots	&	\vdots	&	\vdots	&	\ddots	&	\vdots	&	\vdots	&	\vdots\\
0	&	0	&	0	&	\dots	&	-2	&	1	&	0\\
0	&	0	&	0	&	\dots	&	1	&	-2	&	1\\
1	&	0	&	0	&	\dots	&	0	&	1	&	-2\\
 \end{pmatrix}
 \end{split}
 \label{eq:jacobian_symmetric}
\end{equation}
The eigenvalues of this matrix can be written in an explicit form \cite{Gray2005}:
\begin{equation}
 \lambda_j=-2K\cos\Delta\phi\left(1-\cos\frac{2\pi (j-1)}{N}\right)\,,\qquad j=1,\dots, N\,.
 \label{eq:eigenvalues}
\end{equation}
The expression in the bracket is non-negative, so in order to have a stable equilibrium we need $\cos\Delta\phi>0$, 
which implies:
\begin{equation}
 -\frac{\pi}{2}<\Delta\phi<\frac{\pi}{2}.
\end{equation}
This result can be formulated in terms of the state index:
\begin{equation}
\label{stability}
 -\frac{N}{4}<m<\frac{N}{4}.
\end{equation} 

This is the same results as the one given in \cite{Roy2012888}.


\subsubsection{Case (b) states}
\label{sec:case_b}
Case (b) is present when Eq.~(\ref{eq.cond2}) is true all over the system. Similarly to case (a) we evaluate the Jacobian of the system near the $\mathbf{u}^*=(u_1^*,\dots,u_i^*,\dots)$ equilibrium
point. Following the same argument as before and using Eq.~(\ref{eq.cond.pm_b}) we can write up the Jacobian:
\begin{equation}
\begin{split}
\label{eq.jacobi_case_b}
 \mathbf{J}=K \cos\Delta\phi
 \begin{pmatrix}
0	&	-1	&	0	&	\dots	&	0	&	0	&	+1\\
-1	&	0	&	+1	&	\dots	&	0	&	0	&	0\\
0	&	+1	&	0	&	\dots	&	0	&	0	&	0\\
\vdots	&	\vdots	&	\vdots	&	\ddots	&	\vdots	&	\vdots	&	\vdots\\
0	&	0	&	0	&	\dots	&	0	&	+1	&	0\\
0	&	0	&	0	&	\dots	&	+1	&	0	&	-1\\
+1	&	0	&	0	&	\dots	&	0	&	-1	&	0\\
 \end{pmatrix}
 \end{split}
\end{equation}
Here, however, we do not intend to calculate the eigenvalues explicitly, we only show that the Jacobian must have at least one positive eigenvalue, thus proving the instability of these states.
Having a symmetric matrix indicates that all the eigenvalues are real, $\lambda_i\in\mathbb{R}$. From the trace of the Jacobian we get the sum of the eigenvalues:
\begin{equation}
 \label{eq.trace}
 \sum_{i=1}^{N}J_{ii}=\sum_{i=1}^{N}\frac{\partial \dot{u}_i}{\partial u_i}=\sum_{i=1}^{N}\lambda_i=0.
\end{equation}
Having zero as trace can mean two things: 
\begin{enumerate}[\textbullet]
 \item $\lambda_i=0,\quad\forall\, i \Rightarrow$ neutral linear stability,
 \item $\sum_i \lambda_i^+=\sum_i |\lambda_i^-|\Rightarrow$ instability. 
\end{enumerate}
Here $\lambda_i^+$ denotes the positive, while $\lambda_i^-$ values are the negative eigenvalues.
Now we show that the latter case is true, namely $\mathbf{J}$ has negative and positive eigenvalues, thus these states are unstable.
To demonstrate that,
let us suppose that all $\lambda_i$ eigenvalues of $\mathbf{J}$ are~0. Using spectral decomposition we can write:
\begin{equation}
 \label{eq.spectral}
 \mathbf{J}=\mathbf{S\Lambda S}^{-1},
\end{equation}
where $\mathbf{\Lambda}$ is a diagonal matrix, having the eigenvalues of the Jacobian on the main diagonal, while $\mathbf{S}$ is a square matrix having the corresponding eigenvectors
as columns. On the other hand:
\begin{equation}
 \label{eq.contradiction}
 \mathrm{if}\quad \lambda_i=0,\quad\forall i=1,\ldots N \quad \Rightarrow \quad \mathbf{\Lambda} = \mathbf{0}_N \quad \Rightarrow \quad \mathbf{J} = \mathbf{0}_N.
\end{equation} 
This is clearly a contradiction since in our case the Jacobian is explicitly given in 
Eq.~(\ref{eq.jacobi_case_b}), hence $\mathbf{J}$ must have nonzero eigenvalues as well.
Now as a consequence of Eq.~(\ref{eq.trace}) we can say that these kind of stationary states are unstable, due to existence of positive eigenvalues.
\subsubsection{Case (c) states}
\label{sec:case_c}

For the third branch we also have to rely on the trace of the Jacobian since 
its general form is not clear, thus the explicit form of the eigenvalues are not known. The reason is that these states can be constructed from  any 
combination of the first two cases, which obeys Eq.~(\ref{eq.case_c.sum}).

The trace of $\mathbf{J}$ for the case (c) states is:
\begin{equation}
 \label{eq.trace_c}
 \sum_{i=1}^{N}J_{ii}=\sum_{i=1}^{N}\frac{\partial \dot{u}_i}{\partial u_i}=-2K(2n-N)\cos\Delta\phi,
\end{equation}
where $n$ is defined as in (\ref{eq.case_c.sum}), indicating the number of phase shifts having $\Delta\phi$ value while the other $N-n$ are equal to $(2p+1)\pi-\Delta\phi$.
In addition to the $\mathrm{Tr}\,\mathbf{J}=0$ case discussed before, here other situations can also arise:
\begin{enumerate}[\textbullet]
 \item $\sum_{i=1}^{N}J_{ii}>0\Rightarrow$ unstable state,
  \item $\sum_{i=1}^{N}J_{ii}<0\Rightarrow$ stability can not be determined. 
\end{enumerate}
In conclusion these states are unstable if: 
\begin{enumerate}[\textbullet]
\item  $n\leq N/2$ and $|\Delta\phi|\leq\pi/2$,
\item $n\geq N/2$ and $|\Delta\phi|\geq\pi/2$. 
\end{enumerate}
In any other case the stability could not be determined through linearization, 
however computer simulations suggest that these highly symmetry breaking states are not stable.

\subsection{Computer experiments}

Numerical integration of system (\ref{eq.1}) starting from uniformly distributed initial $\theta_i$ values
will reveal only the stable stationary states. It is worth noting that a uniform distribution of the states in the $\theta$ space will lead to a uniform distribution in the $\Delta \phi$ space as well.

\begin{figure}[t]
  \centering
  \includegraphics[width=0.75\linewidth]{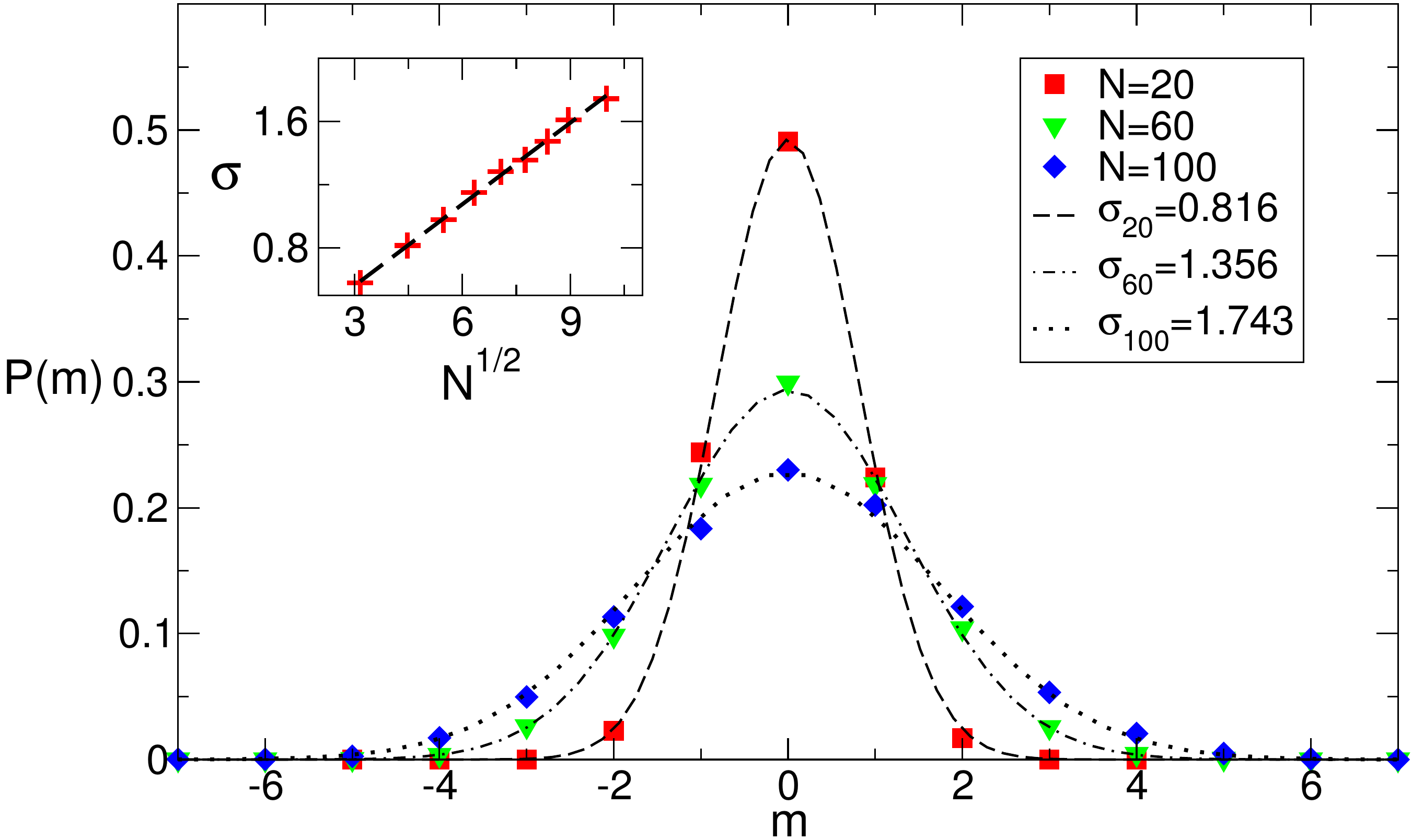}
  \caption{(color online) Probability distribution of the final states for different oscillator numbers $N$, and the normalized Gaussian probability density envelope curve fitting the discrete points. The inset shows the scaling of the standard deviation, $\sigma\propto\sqrt{N}$. The distributions are obtained over 5000 runs for the $K=1.5$, $\omega_0=2$ parameters.}
  \label{fig.1}
\end{figure}

A series of simulations performed on systems with sizes ranging from $N=4$ to $N=100$ confirms the results 
presented in equations (\ref{winding}) and (\ref{stability}). In agreement with the simulation results presented in
\cite{wiley_strogatz_2006} we also find that for  $-\frac{N}{4}<m<\frac{N}{4}$ the probability distribution of the stable states
can be described with a Gaussian envelope curve (Fig.~\ref{fig.1}). As it is visible in the inset of Fig.~\ref{fig.1}, the 
standard deviation of the probability distribution scales linearly with the square-root of the system size, a result emphasized already in  \cite{wiley_strogatz_2006}. We reproduce here this 
result both for showing the validity of our numerical methods and for an easier understanding of the pattern selection procedure. Based on this figure and the Gaussian fit one can offer a first probabilistic prediction for the final stationary state selected by a system with fixed $N$ and $K$ parameters with random initial phases.  
Our computer experiments also shows that the distribution does not change if one changes the coupling strength $K$, which is a trivial consequence of equations (\ref{eq.1}), if one rescales properly the time. 

\section{Dynamics of the system}
\label{sec:dynamics}
\begin{figure}[t]
  \centering
  \includegraphics[width=0.5\linewidth]{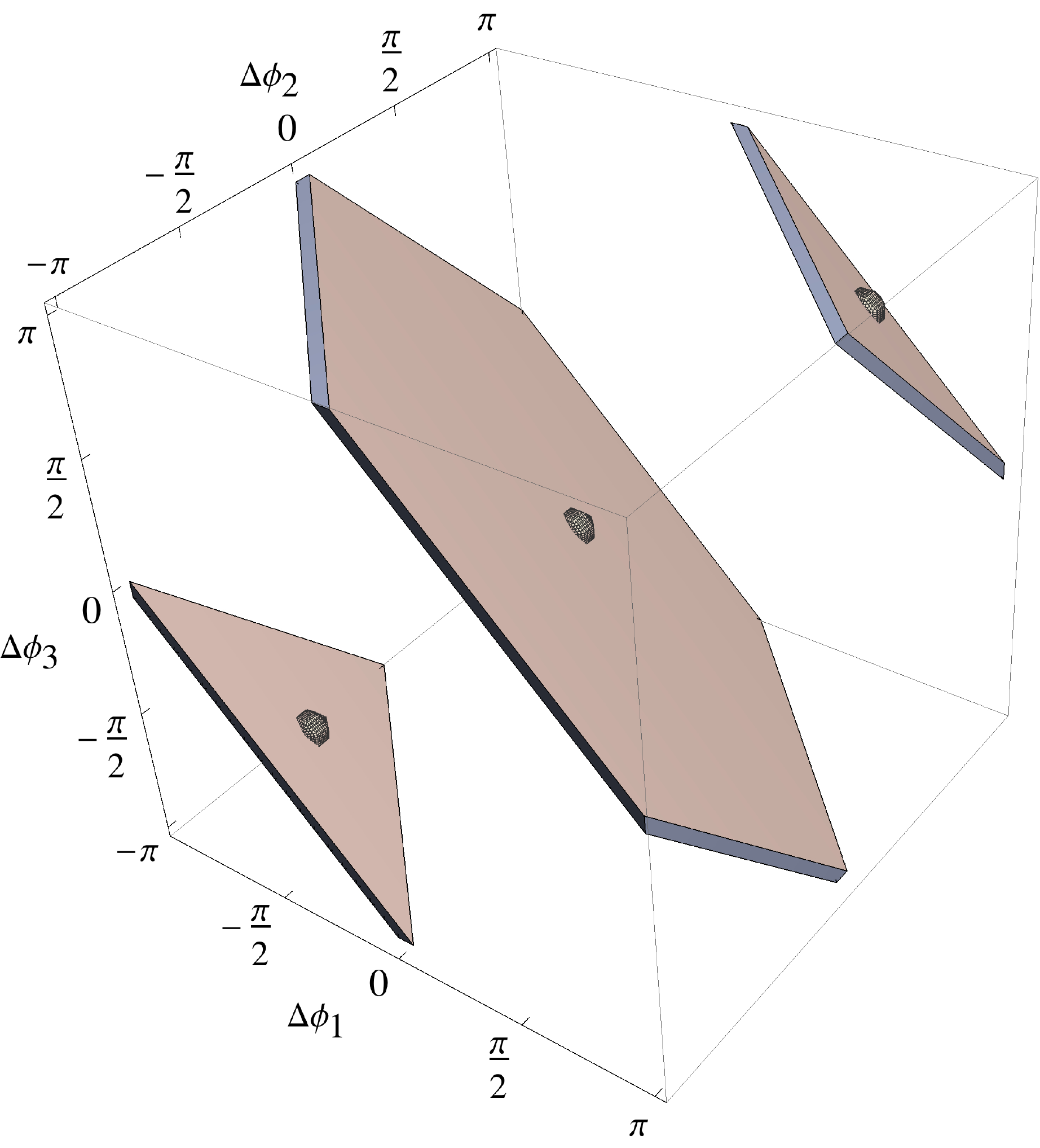}
  \caption{(color online) The planes defined by condition (\ref{winding}) for a system of $N=3$ rotators. Black spheres indicate the allowed stationary states of the system (stable and unstable
  states corresponding to case (a)). The central large plane
  is for $m=0$, while the other two planes correspond to winding numbers $m=\pm 1$.  }
  \label{planes}
\end{figure}

Let us consider first the evolution of the system in the $N$ dimensional $\Delta \phi$-space. 
Since $\Delta \phi_i$ is defined for $-\pi\le  \Delta \phi_i < \pi$ the allowed phase space is confined in a hypercube centered in the origin of the $N$-dimensional
space (for the 3D case see Fig.~\ref{planes}).  In the stationary states all phase shifts are equal, so the attractors lie on the main 
diagonal of the hypercube (large black points in Fig.~\ref{planes}) . We have shown that these states are discrete, thus the attractors represent distinct points on this line.
In Section \ref{sec:overwiev}, Eq.~(\ref{winding}), we have also proved that at each time moment of the dynamics:
\begin{equation}
\sum_{i=1}^N \Delta \phi_i=2m\pi\,,
\label{period}
\end{equation} 
with $-N/2\le  m < N/2$ and $m\in\mathbb{Z}$.
Eq.~(\ref{period}) can be interpreted as the equation of a plane in the $N$-dimensional space, determined by the phase shifts in the system. 
The $N$-dimensional characteristic point of the system can only exist on the planes defined by various $m$ values in Eq. (\ref{period}). These planes are parallel to each other, and  as $m$ increases 
in absolute value the area of the cross-sections of the planes and the hypercube
gets smaller. The 3D case is illustrated in Fig.~\ref{planes}, where the larger central plane is for $m=0$, and the the two
smaller planes are for $m=\pm1$. During the evolution of the system the characteristic point is moving on these planes. 
Jumps between the planes are also possible when the configuration of the phase shifts changes in a way that the winding number defined by the sums (see Eqs.~(\ref{deltaN1}) and (\ref{winding}))
is altered by +1 or -1. This occurs when the characteristic point of the system reaches the boundary of a plane (i.\ e.\ phase shift 
between two oscillators crosses the $\pi$ or -$\pi$ value).
This representation gives a first qualitative image for the dynamics of the oscillator ensemble. 

\begin{figure}
 \centering
 \includegraphics[width=0.75\linewidth]{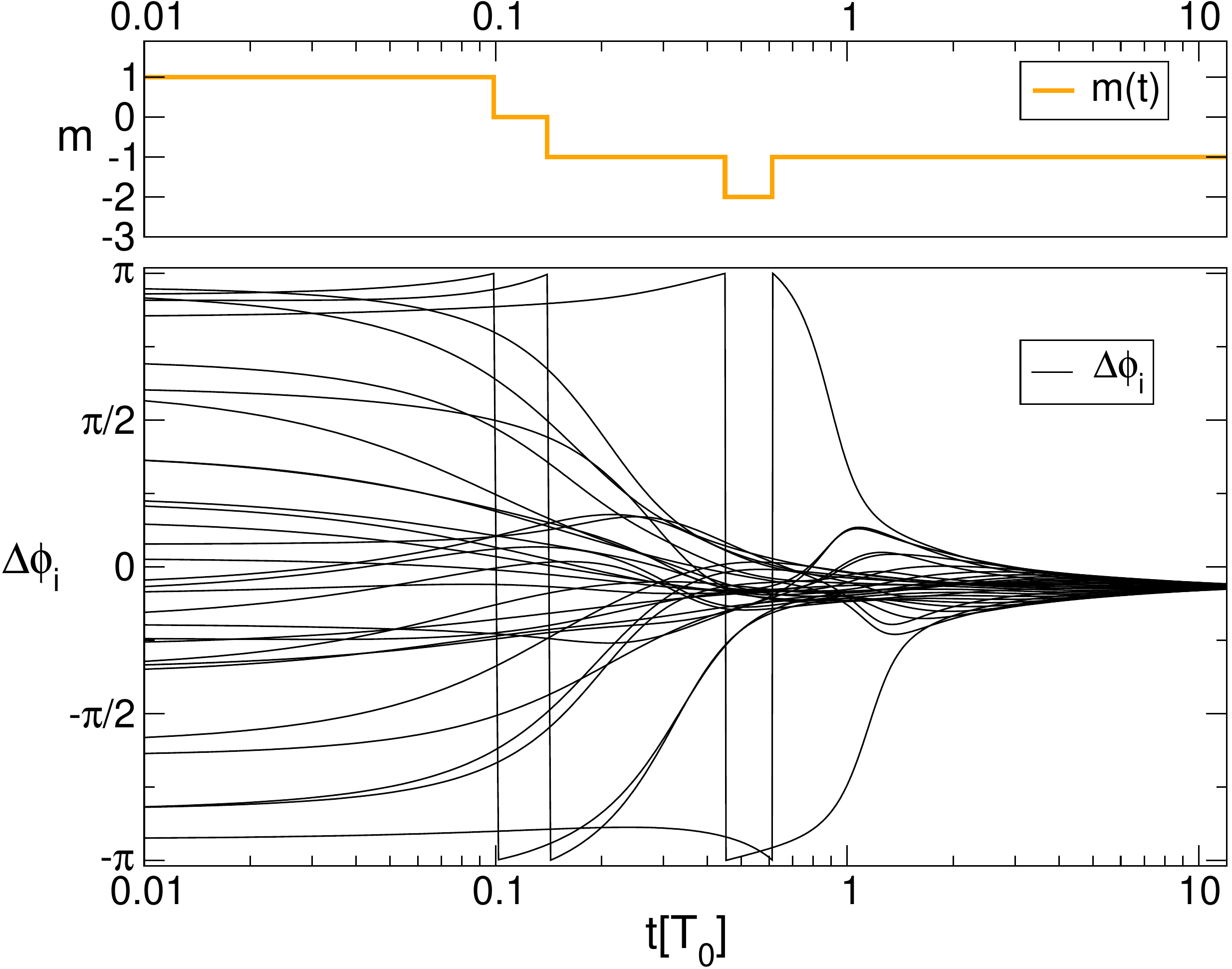}
 \caption{(color online) Top: the $m$ winding number as a function of time. Each integer value corresponds to a plane defined by Eq.~(\ref{period}). Changes in the value indicate the jumps between the planes. Bottom: characteristic time evolution for the phase shifts $\Delta \phi_i$ between neighboring oscillators. 
System parameters: $N=30$, $K=0.75$. Note the logarithmic time-scale.  }
 \label{fig.6}
\end{figure}

For $N>3$ the actual trajectories cannot be easily visualized. In order to get some useful information on the dynamics of the system one option is to  plot all $\Delta \phi_i$ phase shifts as a function of time.
A characteristic time-evolution is sketched in Fig.~\ref{fig.6}. One can observe on the plotted dynamics the jumps that 
occur at the border of the planes, see the steps in the winding number $m$.  This representation suggests already the complexity of
the dynamics, and raises the problem of predicting the final stationary state of the system after a finite 
time in the evolution. Our aim in the following sections is to discuss 
prediction possibilities and analyze their success rate critically.

The predictability of the final stationary state right from the beginning of the dynamics is strongly influenced by the number of oscillators. The complexity of the dynamics is increasing with the system size: 
the volume of the phase space is exponentially growing with $N$
(the allowed hyperplanes defined by Eq.~(\ref{period}) are confined in an $N$-dimensional hypercube) while the number of stable stationary states scales
linearly with the dimensionality of the system. Moreover the increasing variance of the distributions in Fig.~\ref{fig.1} also indicates 
 growing complexity in the system. All of these suggest that predicting the final stationary state from random initial conditions becomes more and more difficult as $N$
is increasing.

\begin{figure}[t]
 \includegraphics[width=\textwidth]{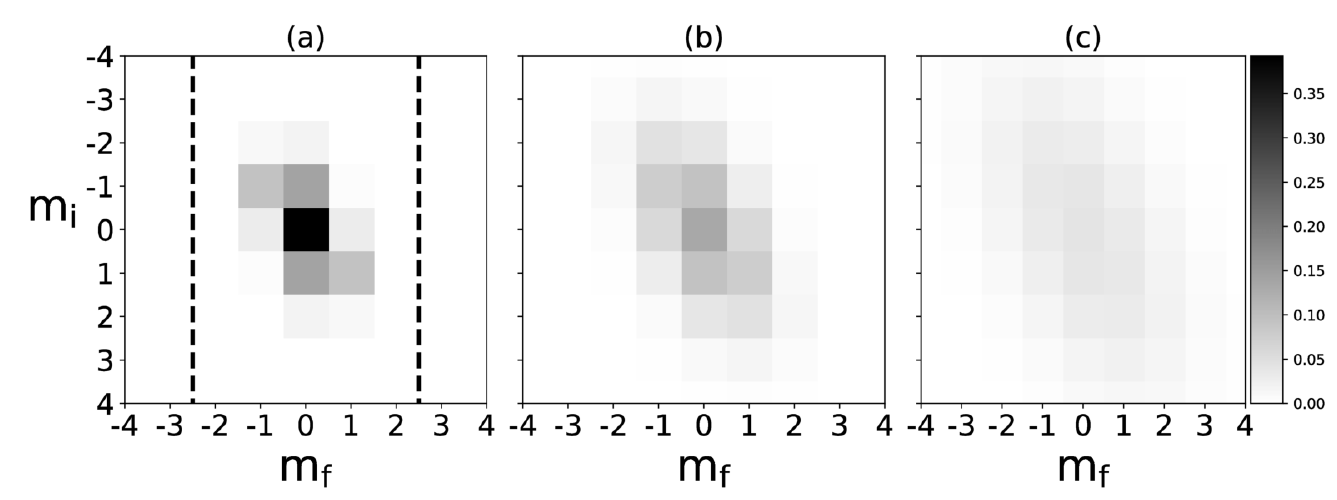}
 \caption{A gray-level code representation for the dynamics of the $m$ winding number.  The $m_i$ values indicate the winding number at $t=0$, while $m_f$ is the winding number in a very late stage (presumably the final stationary state) of the dynamics. Each possible
 $m_i\rightarrow m_f$ transition has some probability which is marked by the gray-level color of the cell as it is illustrated by the attached color-code. The initial state is randomly generated, and we consider 
 increasing number of  oscillators: (a) $N=9$, (b) $N=27$, (c) $N=81$. 
 Since only a fraction of possible $m$ values correspond to stable stationary states (see Eq.~(\ref{stability}) ),
 cells with $|m_f|\geq N/4$ are not possible and we excluded them by the dashed lines so that
 impossible "transitions" lie outside of the dashed lines. }
 \label{fig.matrix}
\end{figure}

 In order to visualize the growing complexity we compare the initial and late stage positions through the $m$ winding number in 
Fig.~\ref{fig.matrix}. We considered random initial phases and we calculated the initial winding number $m_i$. As we let the system to evolve it 
converges to an ordered state with some index $m_f$ (winding number in the late stage of the dynamics, presumably the final state). Arranging these pairs into a matrix we can assign a probability to each "transition" such that these probabilities sum up to 1. The strong peak at 0$\rightarrow$0 transitions for small systems gradually smoothen out as we increase the number of oscillators, indicating that
more transitions are possible, therefore reliable predictions at $t=0$ are difficult to make.

\section{Predicting the final stationary state}
\label{sec5}

It is a natural question now, how and when we are able to identify the final stationary state, if the system is initialized with random phases. 
Similarly to the known Kuramoto order parameter~$r_0$ (for $m=0$), one can define a generalized order parameter 
$r_m\in[0,1]$ for each $|m|>0$ case (a) (see eq. (\ref{eq.4})) stationary state. This parameter will give a useful information on how well the system 
approached the given stationary state with index $m$: 
\begin{equation}
 \label{eq.7}
  r_{m}(t)\mathrm{e}^{i\psi_m(t)}=\frac{1}{N}\sum_{j=1}^{N}\mathrm{e}^{i[\theta_j(t)-(j-1)\frac{2m\pi}{N}]}.
\end{equation}

\begin{figure}[t]
 \centering
 \includegraphics[width=0.75\linewidth]{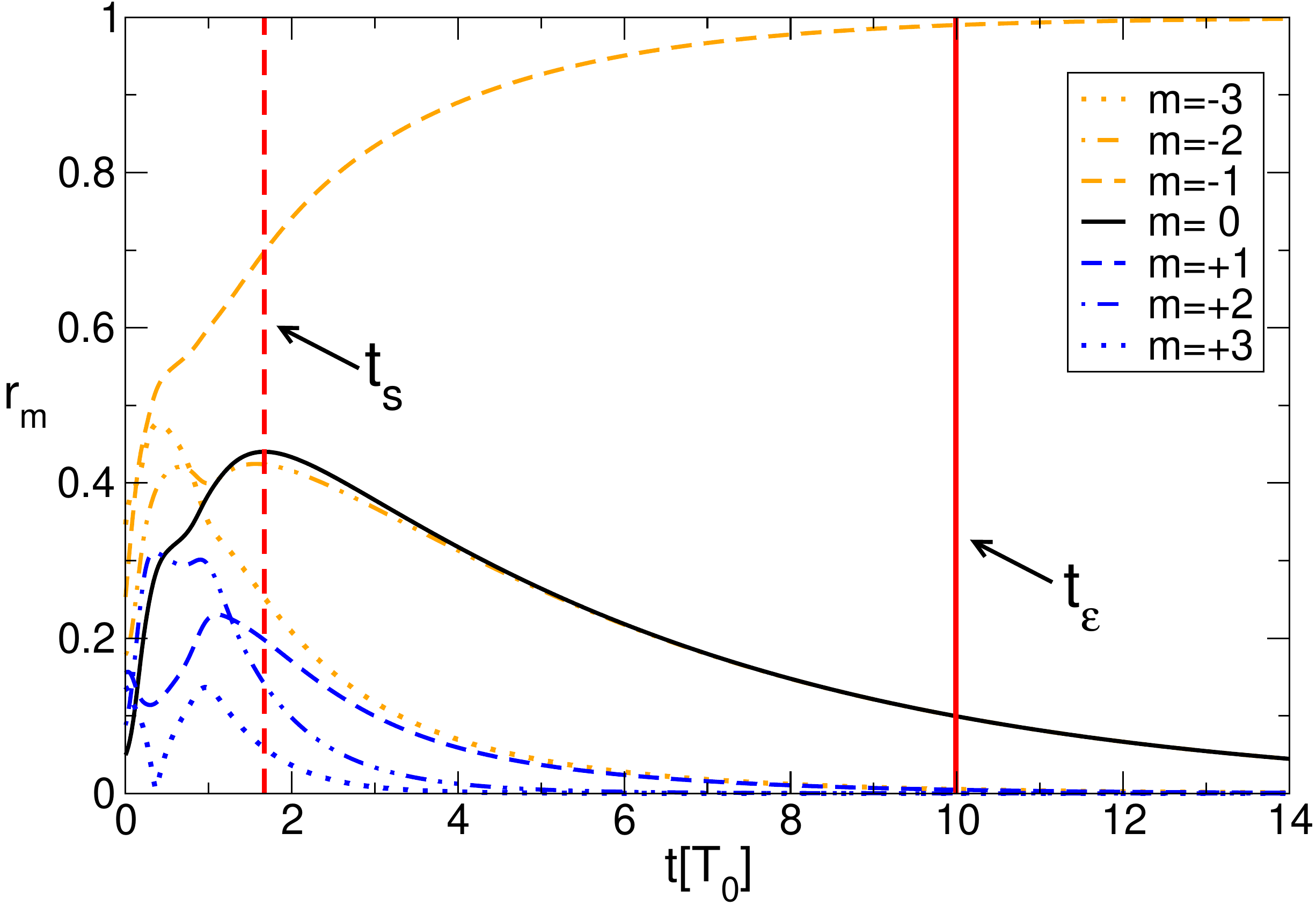}
 \caption{(color online) Characteristic time-evolution for the order parameters $r_m$ of different final states as a function of time, for $|m|\le 2$ to avoid overcrowding. The dashed and continuous vertical lines indicate the $t_s$ and $t_\varepsilon$ time-moments of prediction, where $t_s$ is defined by the derivatives of the $r_m$ order parameters, see Eq.~(\ref{eq.selection}), while $t_\varepsilon$ is determined by the first
 moment of crossing the threshold in Eq.~(\ref{eq.threshold}). The data is obtained during the same run as in Fig.~\ref{fig.6} ($N=30$, $K=0.75$).}
 \label{fig.7}
\end{figure}

Naturally for $r_k=1$, the system is in the stationary state characterized by the index $k$, while all the other ($m\ne k$) order parameters are 0. The smaller $r_m$ is,
the further the system is from the corresponding stationary state. This generalized order parameter is also a useful tool for following the 
time-evolution of the system. Now that we can quantify the level of order in the system a straightforward solution for predicting the final stationary state would be to define an $\varepsilon$ tolerance
value: if one of the order parameters approaches 1 within the given tolerance we say that the state with the specific winding number will be selected:
\begin{equation}
\mathrm{if}\quad r_{m^*}(t)>1-\varepsilon\quad\Rightarrow\quad\lim_{t\rightarrow\infty}r_{m^*}(t)=1.
\label{eq.threshold}
\end{equation}
The quantity $1-\varepsilon$ is also called threshold.
In Appendix A we show that close to the fixpoints corresponding to the 
selected final winding number $m^*$, the order parameter 
$r_{m^*}$ approaches value 1, 
while all the other $r_{m\neq m^*}$ order parameters go to 0.

 Characteristic results for the time evolution of the $r_m$ values are shown in Fig.~\ref{fig.7}. 
As one can see the dynamics is built up by two stages: a first short stage with nontrivial evolution and a second slow exponential relaxation. 
The continuous 
vertical line here denotes a time moment when one of the order parameters reaches a fixed threshold
($\epsilon=0.01$). In this example it is clear that the threshold is set too high,
since the moment of decision is way too ahead in the relaxation stage resulting in waste of CPU time. On the other hand at the cost of time one gains precision since speeding
up too much the algorithm by a lower threshold may end up in wrong predictions. A less time consuming and self-explanatory method would be to identify the end of the first
  stage because the relaxation process holds no new information about the dynamics. Since the order parameters describe how well the system approaches one state (this
  not necessarily mean euclidean distance in the phase space)  the relaxation process can be viewed as the time interval during which only one order parameter is increasing, meaning only one state is approached.
Hence we argue that there is a $t_s$ time moment in the relaxation process  so that if $t\ge t_\mathrm{s}$ only the $r_k(t)$ corresponding to the selected state $k$ keeps increasing:
\begin{equation}
\begin{aligned}
\dot{r}_k(t\ge t_s) &> 0 \\ 
\dot{r}_{j\ne k}(t \ge t_s) &< 0\,.
\end{aligned}
\label{eq.selection}
\end{equation}

A final state prediction method based on this observation would have the strength that it does not need externally set and unknown threshold parameters for prediction and probably it would be also faster as it can be seen in Fig.~\ref{fig.7}, where the dashed line
indicates the moment in which we make decision. 

\begin{figure}[t]
 \centering
 \includegraphics[width=0.75\linewidth]{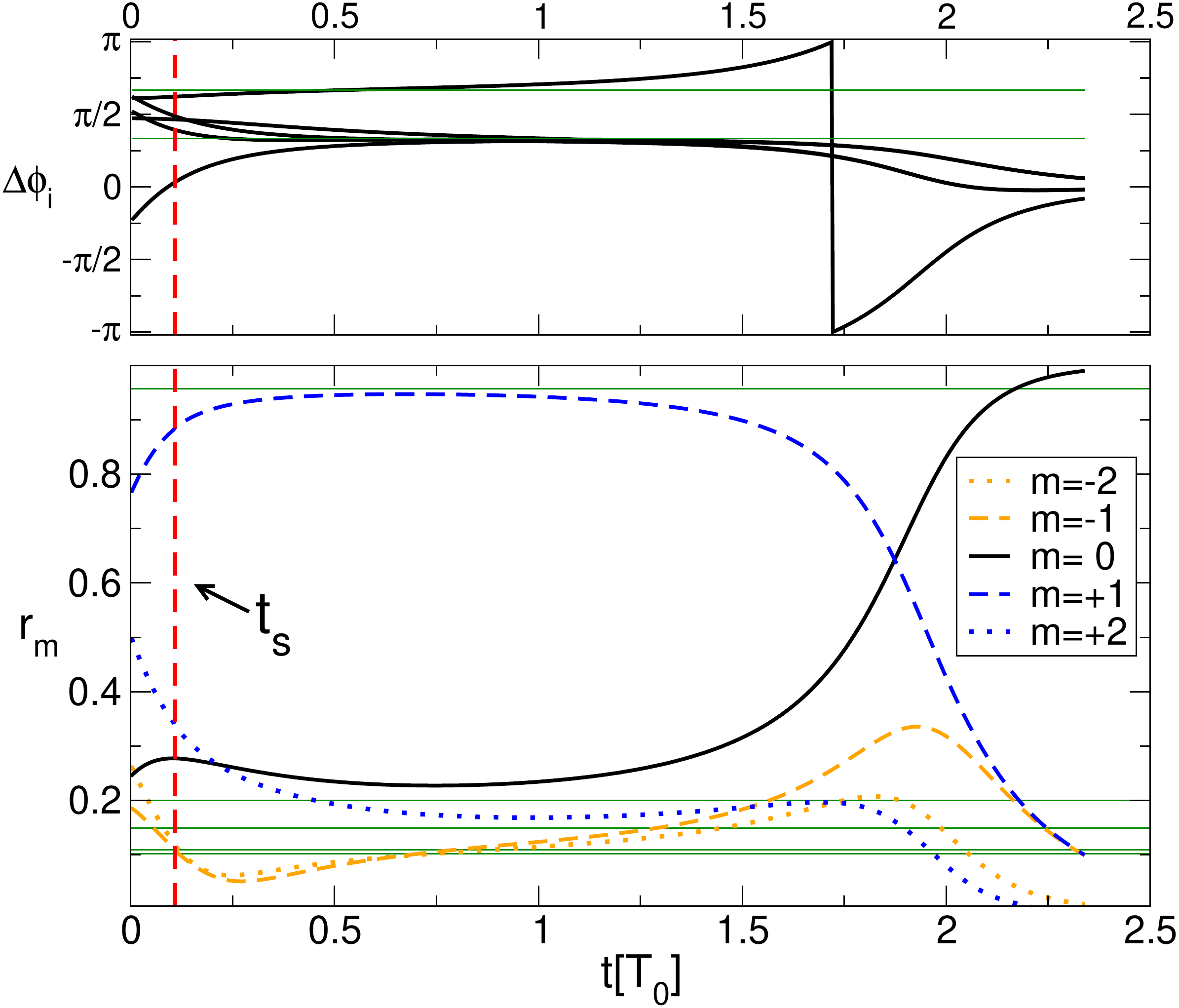}
 \caption{(color online) Top: time evolution of the $\Delta\phi_i$ phase shifts for $N=5$. Thin horizontal lines mark the values of the phase shifts in the saddle point sketched in the right part of Fig.~\ref{fig.case_c}. As the characteristic point passes by the saddle the phase shifts approach the marked values. Bottom: time evolution of the order parameters for the same
 run as in the top panel. This scenario illustrates a case where the prediction based on the derivative of the order parameter fails. The horizontal lines represent the values of order parameters in the saddle point which is approached by the system. Since the $r_{+1}$ parameter has a much larger value in the saddle as
 the others, approaching that point, this order parameter has to grow, while the other ones have to decrease. This leads us to the incorrect prediction that the
 $m=+1$ state is selected (marked by the vertical dashed line), even though the relaxation process has not started yet. In the end it is clear that the $m=0$ state will be preferred.  }
 \label{fig.saddle}
\end{figure}

However it must be noted that the reverse argument may not necessarily be true, namely if only one order parameter is increasing this does not guarantee that the system entered in the relaxation process, which can also lead to incorrect predictions. Simulations show that this can happen due to 
existence of saddle points in the phase space. These saddle points are the cases (b) and (c) stationary points of the potential function described in the previous section \ref{sec:stationary}. 
Some of these points with strong attracting character (saddle points with substantially more negative eigenvalues than positive ones) may distort the trajectories in such a way that the above method based upon the derivatives of the order parameters fails. We demonstrate this by showing a specific example. In the bottom panel of Fig.~\ref{fig.saddle} the time evolution of the order parameters is presented for a system with five oscillators ($N=5$). The continuous horizontal lines mark the values of the order parameters for the state depicted in the right panel of Fig.~\ref{fig.case_c}. Our method
predicts early in the beginning that the  $m=+1$ state will be selected, because all the others are decreasing. This moment is marked by a vertical dashed line. However it is clear that not the $m=+1$ state
is being approached but the saddle point. These points are also ordered states and 
 usually one order parameter stands out from the others. If the system gets close to a saddle point the values of the order parameters start resembling the ones that the unstable stationary points has. Then the order parameters that are smaller than desired would increase and in the meantime the greater ones have to decrease. This may lead to the
case when only one is increasing which results in an incorrect prediction. The upper part of Fig.~\ref{fig.saddle} presents the same scenario on the level of phase shifts. The horizontal lines there indicate the value of the phase shifts in the saddle point in question.

\begin{figure}[t]
 \centering
 \includegraphics[width=0.75\linewidth]{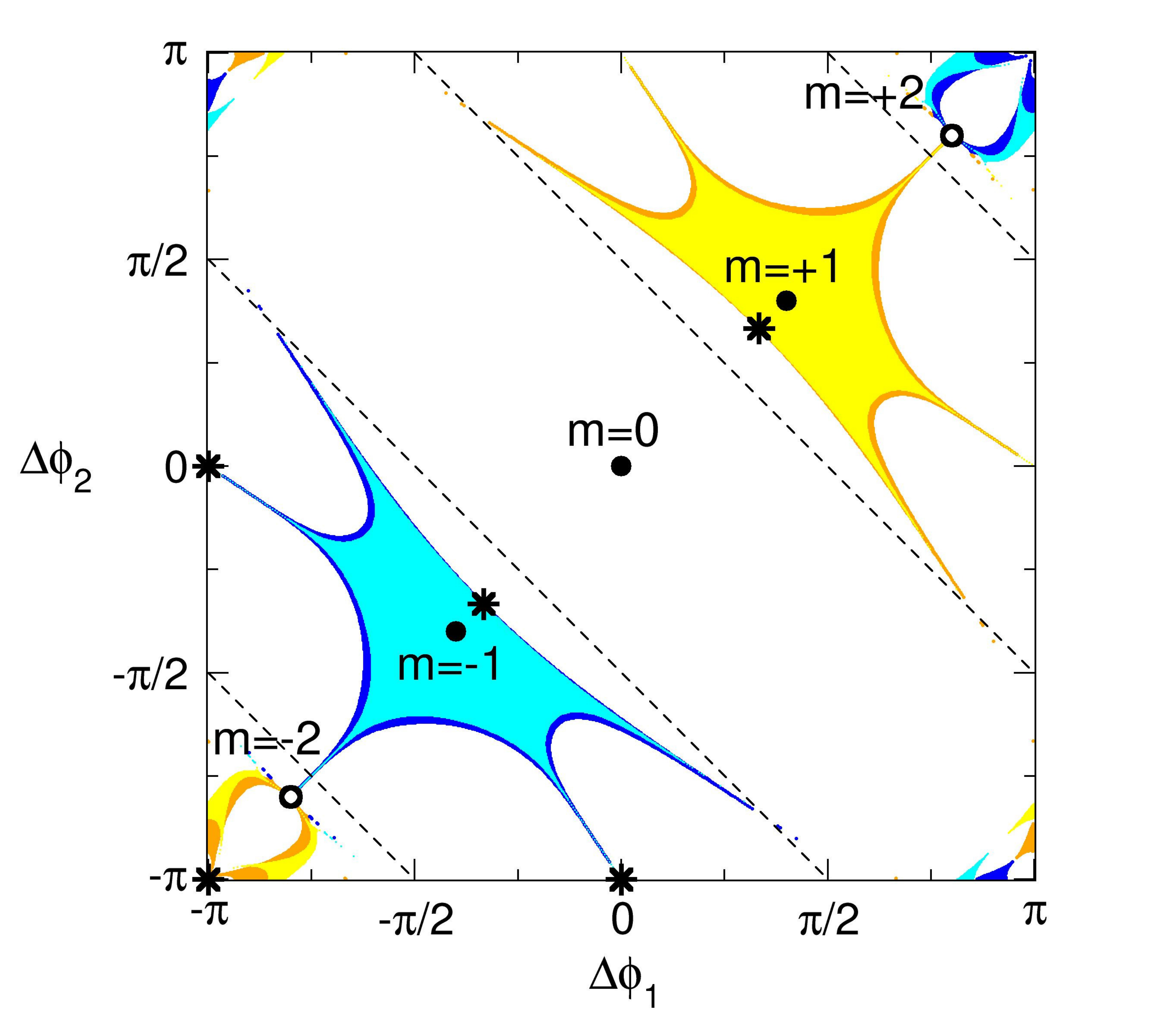}
 \caption{(color online) Cross section of the basins of attraction containing the main diagonal of the $\Delta\phi$ space for a system with $N=5$ oscillators. Black dots indicate the stable fixpoints ($m=\{-1,0,1\}$), open circles represent the unstable states ($m=\pm2$) and the stars are the saddle points. Dashed lines illustrates the borders of different planes. White areas mark the
 basin of attraction of the $m=0$ synchronized state, while the colored areas belong to the basins of $m=\pm1$ states. Brighter regions are obtained by a high  $1-\varepsilon=0.999$ threshold value (practically this means that they are accurate),  while the darker areas indicate where our predictions based on the derivatives of $r_m$ are incorrect. }
 \label{fig.section}
\end{figure}

In order to be more clear and make also connection with the results presented in section \ref{sec:dynamics} we study a cross section of the basins of attraction (Fig.~\ref{fig.section}) for this specific case for $N=5$. The $\Delta\phi$-space of an $N$ dimensional system is always $N-1$ dimensional due to the constraint in Eq.~(\ref{winding}) limiting the dynamics on hyperplanes. 
This means that we are able to construct a two dimensional cross section of the phase space by fixing two $\Delta\phi$ values, varying other two and the last has to be calculated from the condition $2m\pi-\sum_{i=1}^{N-1}\Delta\phi_i$. In order to see all the relevant fixpoints (minima, maxima and saddle points) we constructed an
"oblique" cross section with maximal symmetry properties along the main diagonal as the following:
\begin{equation*}
\begin{aligned}
\Delta\phi_1&,\,\Delta\phi_2 - \mathrm{are~varied},\\
\Delta\phi_3&=\Delta\phi_2, \\
\Delta\phi_5&=\Delta\phi_1, &\\
\Delta\phi_4&=2m\pi-2(\Delta\phi_1+\Delta\phi_2). 
\end{aligned}
\end{equation*}
By considering different points in this cross section as initial states and linking them to the final state they belong (determined by choosing a very small $\epsilon=0.001$ value for the stopping threshold), one is able to visualize sections of the basins of attraction around the stable states.
 
In Fig.~\ref{fig.section} there are actually two sets on the top of each other, each constructed with a different method: one by using the mentioned threshold value to have accurate approach to the 
final stationary state and the other obtained by prediction using the derivatives of the order parameter. The dark colored regions indicate the discrepancy between the two method i.e. where the second approach fails. Not surprisingly at all the errors are along the boundaries of the basins and near the local maxima of the potential which also marks the place where different attraction domains meet. The saddle point between the $m=0$ and $m=+1$ stationary points is the one which is approached by the system in the example given in Fig.~\ref{fig.saddle}. In this example the system was initialized on the $m=+1$ plane. From Fig.~\ref{fig.section} we can conclude that if our starting point was on this section it should be in the dark regions of the $m=+1$ plane (each plane is the area between the boundaries at $\pm\pi$ and the dashed lines). With the order parameters derivative we predicted that it will converge to the very same state but actually ended up in the $m=0$ state.

\begin{figure}[t]
 \centering
 \includegraphics[width=0.75\linewidth]{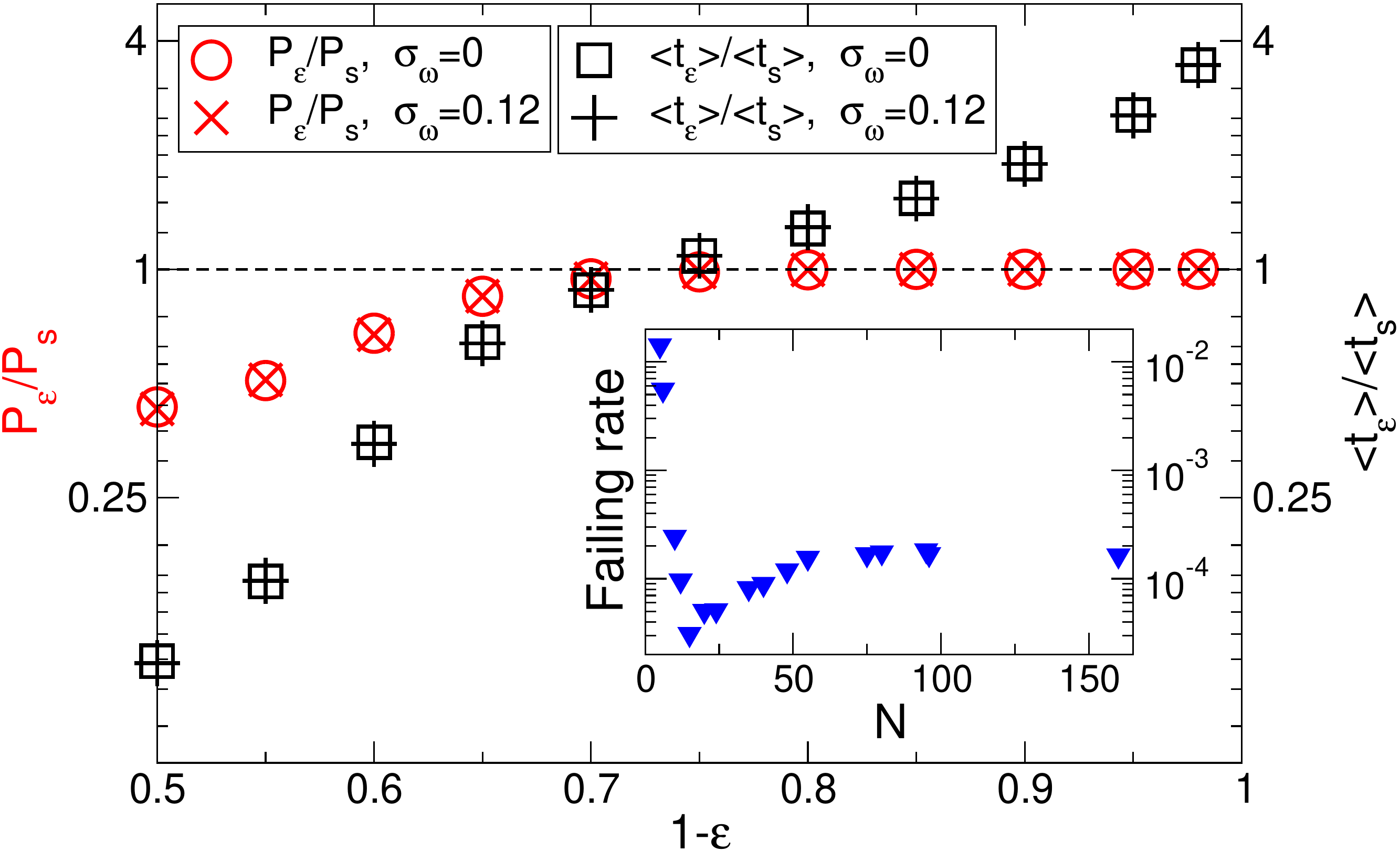}
 \caption{(color online) Comparing the performance of two approaches (fixed threshold value versus the derivatives of the $r_m$ values) for predicting the final state. For different $1-\varepsilon$ threshold values we calculate the 
 the success rate ($P_\varepsilon$) and the average running time ($\langle t_\varepsilon\rangle $) of the first method and compare it to the success rate ($P_s$) and average time ($\langle t_s\rangle$) of the second approach based on the derivatives. For low threshold (high tolerance) the first method is faster albeit its success
rate is poor. At high threshold the success rates are effectively the same, however the second method is more time efficient. The inset shows the failing rate of the second method as a function of system size.
 The success rates are calculated by comparing each to the results evaluated using the high 0.99 threshold ($\varepsilon=0.01$) value. The graph also contains results for systems with nonuniform natural frequencies. $N=18,K=50,K_C=42$. $\sigma_{\omega}$ denotes the standard deviation of the oscillators frequency.}
 \label{fig.precision}
\end{figure}

Seemingly when it comes to final state prediction there is always a trade off between speed and precision as it is illustrated in Fig.~\ref{fig.precision}. The first approach with the proper choice of tolerance can rule out the uncertainties regarding the evolution of the system, however it ceases to be a prediction in the
traditional sense because it only confirms information already clear to the observer. On the other hand the second method is less time consuming yet it comes with inevitable errors due to the presence of saddle points. Its failing rate (inset of Fig.~\ref{fig.precision}) may seem statistically irrelevant but still it is never zero.

One can think now about generalizing the discussion to the case of nonuniform oscillators. An ensemble of nonuniform oscillators with the same coupling topology and arbitrary $\{\omega_i\}$
frequencies can synchronize in state with winding number $m$ only when $K>K_c^{(m)}$ ($-N/4<m<N/4$) \cite{ochab} (for details on $K_c$ please consult \ref{sec:appendix_B}). Data presented on Fig.~\ref{fig.precision} indicates
that for $K>K_c^{(\pm N/4)}$ the dynamics and the statistics show no significant alterations from the homogeneous case. A detailed study of the dynamics is however  much more complicated and we do not consider it in the present study.  A first reason for the increased difficulty is that for not too large ensembles (similar sizes as the one used in the case of homogeneous systems) the specific $\omega_i$ frequency values and their order also influences the critical $K_c^{(m)}$  values. Therefore, it is not enough to fix only the parameters of the distribution function, one needs the specific realization of the~$\omega_i$ frequency values drawn from this distribution, which will influence directly the synchronization properties and largely increases the parameter space of the system. Moreover having non-uniform
natural frequencies will further complicate the potential function in Eq.~(\ref{eq.potential}) with $N$ additional $\omega_iu_i$ terms which changes the easily interpretable form of the stationary states. Results show that the dynamics of the homogeneous system is already rich and worth to study, and the agreement between the two cases hints that the key phenomena are the same. This suggests that such generalization might not reveal as many new information about the dynamics while the derivation becomes less simple and clear.

\section{Conclusion}
Collective oscillation modes were investigated in a ring of identical and locally coupled Kuramoto rotators.  Known results were reproduced by using a different theoretical framework. We identified all possible stationary states including a new class of unstable fixpoints. We proposed a simple algorithm for predicting the final state of the system. A thorough investigations of the method showed however that due to the presence of the unstable fixponts, prediction has always a non-zero failing rate. 
We also found that the complexity of the dynamics increases with the system ($N$) size. Basins of attractions have complicated shapes in the $N$ dimensional state-space and the number of stable stationary states is growing linearly with the system size. Starting the dynamics from a randomly initialized state-point it gets computationally more and more demanding to foresee the final state when $N$ is increased.

\section*{Acknowledgment}

Work supported from the Romanian UEFISCDI grant nr.  PN-III-P4-PCE-2016-0363.
 
\section*{References}
\bibliographystyle{elsarticle-num}
\bibliography{article}

\appendix

\section{}
\label{sec:appendix_A}

In Sec.~\ref{sec5} we introduced the generalized order parameters $r_m$, 
as defined by (\ref{eq.7}), corresponding to different winding numbers $m$ 
in \textbf{case (a)} solutions.
Here we show that close to the fixpoints corresponding to the 
selected final winding number $m^\star$, the order parameter 
$r_{m^*}$ approaches value 1, 
while all the other $r_{m\neq m^\star}$ order parameters go to 0.

Linearizing system (\ref{eq.stab.2}) around the fixpoint 
$\mathbf{u}^*$ for winding number $m^\star$ the solution 
$\mathbf{u}(t)=(u_1,\dots,u_j,\dots)$ may be written as
\begin{equation}
 u_j=\sum_{l=1}^{N}c_l\mathrm{e}^{\lambda_lt}v_{lj}+(j-1)\frac{2m^\star\pi}{N},
 \label{eq:lin_solution}
\end{equation}
where the $c_l$-constants are determined by the initial conditions.
$v_{lj}$ is $j$-th component of the $l$-th eigenvector 
corresponding to $\lambda_l\in\mathbb{R}$ eigenvalue of the symmetric 
Jacobian matrix $J_{ij}$, defined by Eq.~(\ref{eq:jacobian_symmetric}). 
For stable fixpoints $\lambda_{l>1}$ are negative, 
compare Eq.~(\ref{eq:eigenvalues}), and $\lambda_1=0$ 
corresponding to eigenvector $\mathbf{v}_1=(1,\dots,1)$,
as a consequence of the rotational symmetry of the system \cite{Roy2012888}.

The order parameter describing state with winding number $m$
can also be expressed in terms of the new variables~$u_j$,
compare Eqs.~(\ref{eq.7}) and (\ref{eq:variables_u}):
\begin{equation}
 \begin{aligned}
 |r_m|^2 &= \frac{1}{N^2}\sum_j\sum_k\mathrm{e}^{\mathrm{i}(u_j-u_k)-\mathrm{i}(j-k)\frac{2\pi m}{N}}\\
		 &= \frac{1}{N^2}\sum_j\sum_k\mathrm\cos\left((u_j-u_k)-(j-k)\frac{2\pi m}{N}\right)\,,
 \end{aligned}
\end{equation}
where we have used that the exponent is antisymmetric 
with respect to indices $j$ and $k$.

Considering now the order parameter corresponding to the linearized fixpoint,
i.\ e.\ $m=m^\star$, and substituting the solution~(\ref{eq:lin_solution}) 
into the expression above yields:
\begin{equation}
 |r_{m^\star}|^2=\frac{1}{N^2}\sum_{j,k}\cos\left(\sum_lc_l\mathrm{e}^{\lambda_lt}(v_{lj}-v_{lk})\right)\,.
\end{equation}
Note that $v_{lj}-v_{lk}=0$ for $l=1$.
Therefore, for stable fixpoints,
\begin{equation}
 \lim_{t\rightarrow\infty}|r_{m^\star}|^2=\frac{1}{N^2}N^2=1\,.
\end{equation}

On the other hand, for all the other order parameters 
with $m=m^\star\pm s$ where $s>0$, it can be shown that:
\begin{equation}
 |r_{m^*\pm s}|^2=\frac{1}{N^2}\sum_{j,k}\cos\left(\sum_lc_l\mathrm{e}^{\lambda_lt}(v_{lj}-v_{lk})\mp\frac{2\pi s}{N}(j-k)\right).
\end{equation}

In the long term limit hence
\begin{equation}
 \lim_{t\rightarrow\infty}|r_{m^*\pm s}|^2=\frac{1}{N^2}\sum_{j, k}\cos\left(\mp\frac{2\pi s}{N}(j-k)\right)=0\,.
\end{equation}

\section{}
\label{sec:appendix_B}

According to \cite{ochab} $K_c$ is the lowest $K$ value for which  one of the following points ($I_1$ or $I_2$) reaches the value of $2m\pi$: 
\begin{eqnarray}
I_1=  \sum_{i=0}^{N-1}\arcsin\left(\frac{2}{K}\sum_{j=1}^{i}\Delta_j-\Delta_{\mathrm{min}}\right)  \nonumber \\  
I_2=  \sum_{i=0}^{N-1}\arcsin\left(\frac{2}{K}\sum_{j=1}^{i}\Delta_j-\Delta_{\mathrm{max}}\right)     
\end{eqnarray}
We have used here the following notations:
\begin{eqnarray}
\Delta_j&=&\omega_j-\Omega \nonumber \\
\Omega&=&\frac{1}{N}\sum_{i=1}^N \omega_i \nonumber \\
\Delta_{\mathrm{min}}&=&\frac{2}{k}\min_{\{i=0,1,...,N-1\}}\left(\sum_{j=1}^i\Delta_j \right)+1 \nonumber \\
\Delta_{\mathrm{max}}&=&\frac{2}{k}\max_{\{i=0,1,...,N-1\}}\left(\sum_{j=1}^i\Delta_j \right)-1
\end{eqnarray}

\end{document}